\documentclass[conference]{IEEEtran}
\usepackage{cite}
\usepackage{graphicx,epstopdf,url}
\usepackage{subfigure}
\usepackage{verbatim}
\usepackage{textcomp}
\usepackage{slashbox}
\usepackage{amsmath}
\usepackage{amssymb}
\usepackage{amsmath}

\usepackage{multirow}
\usepackage{booktabs}
\usepackage[utf8]{inputenc}  
\usepackage[lined,ruled,boxed,commentsnumbered,linesnumbered]{algorithm2e}
\usepackage{indentfirst}
\usepackage[all]{xy}
\usepackage{dsfont}
\DeclareMathOperator{\EX}{\mathbb{E}}

\newcommand*{\prob}{\mathsf{P}}

\usepackage{amsthm}
\usepackage{mathtools}

\usepackage{array}
\newcolumntype{L}[1]{>{\raggedright\let\newline\\\arraybackslash\hspace{0pt}}m{#1}}
\newcolumntype{C}[1]{>{\centering\let\newline\\\arraybackslash\hspace{0pt}}m{#1}}
\newcolumntype{R}[1]{>{\raggedleft\let\newline\\\arraybackslash\hspace{0pt}}m{#1}}

\begin{document}
\title{Inference Time Optimization \\ Using BranchyNet Partitioning}

\author{
    \IEEEauthorblockN{Roberto G. Pacheco \IEEEauthorrefmark{1} and Rodrigo S. Couto\IEEEauthorrefmark{1}}
   \IEEEauthorblockA{\IEEEauthorrefmark{1}
    Universidade Federal do Rio de Janeiro - PEE/COPPE/GTA - Rio de Janeiro, RJ, Brazil\\
    Email: pacheco@gta.ufrj.br, rodrigo@gta.ufrj.br}
}
\maketitle

\begin{abstract}
Deep Neural Network (DNN) inference requires high computation power, which generally involves a cloud infrastructure. However, sending raw data to the cloud can increase the inference time due to the communication delay. To reduce this delay, the first DNN layers can be executed at an edge infrastructure and the remaining ones at the cloud. Depending on which layers are processed at the edge, the amount of data can be highly reduced. However, executing layers at the edge can increase the processing delay. A partitioning problem tries to address this trade-off, choosing the set of layers to be executed at the edge to minimize the inference time. In this work, we address the problem of partitioning a BranchyNet, which is a DNN type where the inference can stop at the middle layers. We show that this partitioning can be treated as the shortest path problem, and thus solved in polynomial time.
\footnote{\textcopyright2020 IEEE. Personal use of this material is permitted. Permission from IEEE must obtained for all other uses, in any current or future media, including reprinting/republishing this material for advertising or promotional purposes, creating new collective works, for resale or redistribution to servers or lists, or reuse of any copyrighted component of this work in other works.} 
\end{abstract}

\section{Introduction}
\label{sec:intro}
Deep Neural Networks (DNN) are largely employed in machine learning applications, such as computer vision and speech recognition. 
DNN is composed of neuron layers, where each neuron receives inputs and generates a non-linear output. In summary, a DNN architecture is composed of an input layer, a sequence of middle layers, and an output layer.
For image classification, DNN inference executes a feed-forward algorithm to label an input image into one of the predefined classes. 
In this algorithm, each layer receives the output data from the prior layer, executes a computation, and then propagates its output data to the next layer. DNN inference executes this algorithm from the input layer through the middle layers, until it reaches the output layer, which generates the probability for each predefined class~\cite{lecun2015deep}. 

Traditionally, DNNs can be deployed on end devices (e.g., smartphones and personal assistants) or on a cloud server~\cite{satyanarayanan2017emergence,kang2017neurosurgeon}. DNN inference generally requires high computational power, and its execution on resource-constrained end devices can result in a prohibitive processing delay. DNN inference can thus be executed in a cloud computing infrastructure, which is generally equipped with computational resources to accelerate the processing, such as GPUs (Graphics Processing Unit). In cloud-based solutions, end devices gather the data and transmit it to the cloud server, which executes DNN inference. This adds a data communication delay, which is affected by the network behavior between the end device and the cloud, increasing the inference time. This is a severe problem since recent DNN applications, such as cognitive assistance and intelligent vehicles, require high responsiveness~\cite{biookaghazadeh2018fpgas}.
It is necessary to reduce communication and processing delays to achieve high responsiveness. The former is the time required to send data, through the Internet, from the end device to the cloud server. The second one is the time to perform the inference itself, related to the employed hardware. Edge computing emerges as an alternative to reduce the communication delay imposed by cloud computing~\cite{satyanarayanan2017emergence}. This paradigm consists of deploying computational resources at the edge of the Internet (i.e., close to end devices), reducing the communication delay. Edge servers can be installed in locations such as cellular base stations and Wi-Fi access points. 
Nevertheless, edge devices provide a computational capacity significantly lower than the cloud, which adds processing delay. 
Therefore, edge computing reduces communication delay but increases the processing delay as compared to the cloud. Thus, considering responsiveness, there is a clear trade-off between the communication delay and the processing one.

In literature, there are proposals to handle each one of the delays related to DNN inference. To reduce processing delay, BranchyNet proposes classifying an input sample at the middle layers if a certain confidence level is achieved~\cite{teerapittayanon2016branchynet}. 
Regarding communication delay, DNN partitioning proposes to compute the first DNN layers at the edge device and the other ones at the cloud server. 
This proposal is based on the fact that the communication delay to send data from middle layers is significantly lower than the delay to send a raw image~\cite{hu2019dynamic}. This paper combines BranchyNet and DNN partitioning to evaluate the trade-off between processing and communication delay.
To address this trade-off, this paper formalizes an optimization problem whose objective is to find an optimal partition that minimizes the inference time for a BranchyNet. This optimization problem depends not only on network bandwidth and the computational power of edge and cloud, but also aspects inherent to input data, such as image quality. To this end, we model the inference time for BranchyNet. 
Then, to minimize the inference time, we show the equivalency between BranchyNet partitioning and the shortest path problem. Thus, we can derive a globally optimal solution in polynomial-time.

This paper is organized as follows. Section~\ref{sec:related_works} reviews related works about DNN partitioning. Section~\ref{sec:branchyNet} presents background of BranchyNet. Section~\ref{sec:theorical_analysis} models the inference time for this DNN type. Then, Section~\ref{sec:optimization_formulation} formalized the BranchyNet partitioning problem. The experiments are shown in Section~\ref{sec:experiments}. Finally, Section~\ref{sec:conclusion} concludes this paper.


\section{Related Work}
\label{sec:related_works}
To accelerate DNN inference, prior works study how to partition a DNN between edge devices and the cloud server.
Neurosurgeon~\cite{kang2017neurosurgeon} constructs performance prediction models for DNNs. 
It allows to estimate the processing delay at the edge device and the cloud server. Then, these prediction models are combined with wireless network conditions to dynamically select the best partition.
However, Neurosurgeon is limited to chain-topology DNNs. To address this limitation, DADS~\cite{hu2019dynamic} (Dynamic Adaptive DNN Surgery) optimally partitions 
a general DAG (Directed Acyclic Graph) topology DNN.
To this end, DADS treats the partitioning problem as a min-cut problem.
These papers propose DNN partitioning methods, considering DNNs with no side branches. 

Li et al~\cite{li2019edge} propose a partitioning method to BranchyNets~\cite{teerapittayanon2016branchynet} framework. However, different from our work, they use the Branchynet framework only for choosing the DNN size. Instead of using a confidence level threshold in each side branch, their proposal uses a brute force search to choose the branch and the partition decision that achieves a given latency requirement and maximizes the inference accuracy. This method may be unfeasible for increasingly deeper DNNs. We thus assume that confidence level thresholds are well-chosen before the execution of our partitioning method, guaranteeing a high accuracy level. Hence, given the BranchyNet architecture with its thresholds, our paper optimally partitions a BranchyNet to minimize inference time. Moreover, this paper is the first work to model the inference time for a BranchyNet, considering the probability that a sample is classified
at the side branch also as a factor that impacts the inference time. Then, we convert the BranchyNet partitioning problem into a shortest path problem, which can be feasible to increasingly deeper DNNs. 

\section{BranchyNet}
\label{sec:branchyNet}
BranchyNet is a DNN architecture, whose goal is to accelerate inference. This architecture is based on the idea that features extracted in the first layers can label correctly a large number of samples on a dataset. To this end, BranchyNet proposes to modify an original DNN architecture, inserting side branches at the middle layers. These side branches allow an input sample to be classified at middle layers, instead of the output one as in regular DNNs. BranchyNet can use entropy as an uncertainty metric to compute the confidence level of sample classification, to decide if the inference can stop or not at the middle layers~\cite{teerapittayanon2016branchynet}. 

\begin{figure}[!htp]
    \centering
    \includegraphics[width=\linewidth]{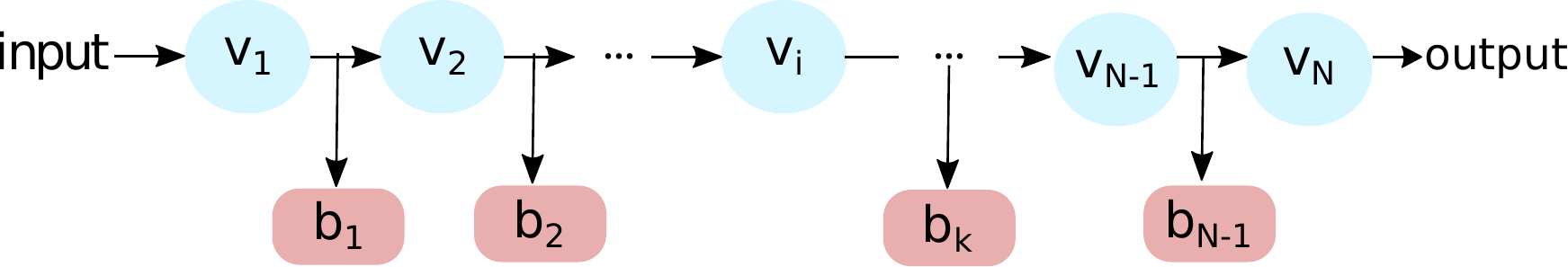}
    \caption{An illustration of a general BranchyNet.}
    \label{fig:branchyNet}
\end{figure}

Figure~\ref{fig:branchyNet} illustrates a generic BranchyNet with $N-1$ side branches. In this figure, the nodes $v_{1}$ to $v_{N}$ represent layers of the main branch, and $b_{1}$ to $b_{N - 1}$ refers to the side branches inserted at those middle layers. In summary, these layers can be of three types: convolutional (conv), max-pooling (max-pool) and fully-connected (fc).
The convolutional layers consist of a set of filters, whose components are learnable parameters during the training process. Each filter is responsible to generate a set of output features using convolutional operations. The max-pooling layers provide robustness to noise in output features of a convolutional layer. To this end, max-pooling layers get the maximum value of a predefined window. 
The output fully-connected layer receives the features extracted by the previous convolutional layers and generates a probability vector, containing the probability that a sample belongs to each predefined class.

Once trained, BranchyNet receives an image, which is processed, layer by layer, until a side branch is reached. 
Then, on the side branch, it computes the confidence level of sample classification based on the probability vector and verifies if this confidence level is less than a threshold. If so, the inference finishes and the class inferred is the class with the highest probability. Hence, this sample is not processed by any next layer, reducing the number of processed layers and thus the processing delay.
Otherwise, the sample is processed by the next layers of the main branch until the next side branch is reached. Then, the whole procedure is performed for this branch. If the sample is not classified in any branch, the inference ends when the output layer is reached. 

When the majority of samples on a dataset cannot be classified at side branches, executing BranchyNet inference at an edge device can introduce processing delay. To avoid that, it is necessary to determine which layer the DNN partitioning must occur to minimize inference time. This decision should take into account the probability of classifying at side branches, the network conditions, and the processing capacity of edge and cloud hardware. Therefore, this work formalizes a BranchyNet partitioning problem, giving the model defined next. 

\section{Partitioning Model}
\label{sec:theorical_analysis}
In this section, we model the inference time for BranchyNet partitioning. To this end, we represent BranchyNet as a graph and define the BranchyNet partitioning problem.

\subsection{BranchyNet Graph}
\label{subsec:graph_branchyNet}
A DNN can be modeled as a DAG $\mathcal{G} = (\mathcal{V}, \mathcal{E})$.
The set $\mathcal{V}$ contains vertices of $\mathcal{G}$. Each vertex $v_{1},\cdots,v_{|\mathcal{V}|}$ represents the layers in a DNN. For instance, the vertices $v_{1}$ and $v_{|\mathcal{V}|}$ represent the input and output layers, respectively. 
The set $\mathcal{E} = \{e_{ij} = (v_{i}, v_{j}) | v_{i}, v_{j}\in\mathcal{V}\}$ contains the links\footnote{In this paper, we use the word ``link'' to denote an edge in a graph, to avoid misunderstanding with edge computing.} of the graph. 
The link $(v_{i}, v_{j})\in\mathcal{E}$ exists if and only if the output data of layer $v_{i}$ feeds the input of the layer 
$v_{j}$. Thus, the layer $v_{i}$ is processed before $v_{j}$.

A BranchyNet can be modeled as a DAG since it is a DNN.
According to Figure~\ref{fig:branchyNet}, a BranchyNet is characterized by inserting a side branch $b_{i}\in\mathcal{B}$ between the middle layers of the main branch, where $\mathcal{B}$ is the set of side branches.
Therefore, we can summarize the BranchyNet architecture into two components: the main branch and side branches.
In this paper, the main branch is modeled as a chain graph, denoted by $\mathcal{P}_{|\mathcal{V'}|}$, where the sub-index $|\mathcal{V'}|$ indicates the number of vertices in $\mathcal{P}_{|\mathcal{V'}|}$. 
In a chain graph, each vertex $v_{i}$ has only one outgoing link, representing its connection to $v_{i+1}$, for all $v_{i}\in\mathcal{V'}$, since $(v_{i},v_{i+1})\in\mathcal{E}$. 
Let $\mathcal{G}_{\text{BDNN}}$ be a DAG of BranchyNet. 
To model the graph $\mathcal{G}_{\text{BDNN}}$, we introduce the vertices $b_{i}\in\mathcal{B}$ into the graph $\mathcal{P}_{|\mathcal{V'}|}$ of the main branch.
In addition, we replace the link
$(v_{i}, v_{i+1})$ with the links $(v_{i}, b_{i})$ and $(b_{i}, v_{i+1})$. In other words, we replace the outgoing link from vertex $v_{i}$ to its neighbor $v_{i+1}$ with a link from $v_{i}$ to the side branch vertex $b_{i}$, and we add a link from $b_{i}$ to $v_{i+1}$.
Thus, at this stage, a BranchyNet also can be modeled by a chain graph denoted by $\mathcal{P}_{|\mathcal{V'}\cup\mathcal{B}|}$. 

\begin{figure*}[!ht]
\subfigure[Edge-only processing.]{\label{fig:only_edge_processing}\includegraphics[width=0.2\linewidth]{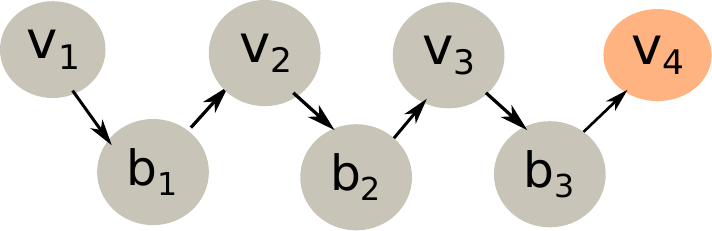}}
\hspace{10mm}
\subfigure[Cloud-only processing.]{\label{fig:only_cloud_processing}\includegraphics[width=0.2\linewidth]{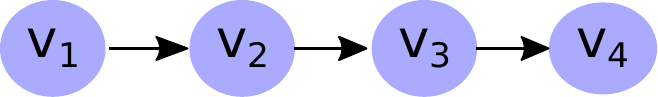}}
\hspace{10mm}
\subfigure[Processing with partitioning]{\label{fig:partitioning_processing}\includegraphics[width=0.2\linewidth]{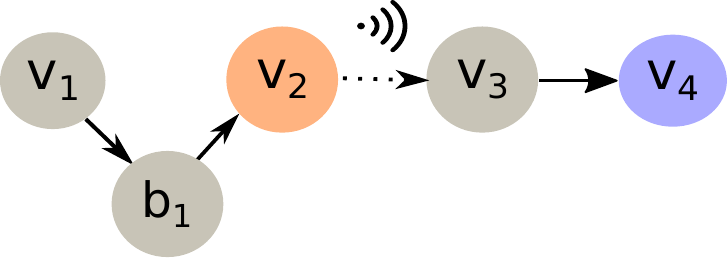}}
\centering
\subfigure{\label{fig:legend_processing}\includegraphics[width=0.7\linewidth]{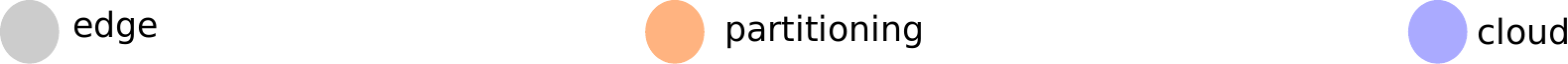}}
\caption{Possible scenarios for DNN processing.}
\label{fig:partitioning_scenarios}
\end{figure*}

\subsection{BranchyNet Partitioning}
\label{subsec:partitioning_branchyNet}
BranchyNet partitioning problem consists of choosing which layer sends its output data from an edge device to the cloud. In this section, we approach this problem as a graph partitioning problem. Given a $\mathcal{G} = (\mathcal{V}, \mathcal{E})$ graph and a positive integer $K$, the graph partitioning finds vertex sets $\mathcal{V}_{1},\cdots,\mathcal{V}_{K}$, such that $\bigcup_{i=1}^{K} \mathcal{V}_{i} = \mathcal{V}$ and$\bigcap_{i=1}^{K}\mathcal{V}_{i}=\emptyset$, which means that each node$v_{i}\in\mathcal{V}$ belongs to only one of these subsets.

BranchyNet partitioning splits the set $\mathcal{V}$ of a BranchyNet graph $\mathcal{G}_{\text{BDNN}}$
into two (i.e., $K=2$) disjoint subsets $\mathcal{V}_{e}$ and $\mathcal{V}_{c}$.
The vertices $v_{1}, \cdots, v_{|\mathcal{V}_{e}|}$ represent BranchyNet layers processed at the edge. The vertices $v_{i}\in\mathcal{V}_{c}$ are the layers processed in the cloud. 
As in graph partitioning, each vertex belongs to only one subset thus the layer $v_{i}$ is processed only at the edge device or at the cloud server, which means that
$\mathcal{V}_{e}\cap\mathcal{V}_{c}=\emptyset$. 
As explained in Section~\ref{subsec:graph_branchyNet}, we model the main branch of Branchynet as a chain graph.
Thereby, formally, the partitioning task must find a layer $v_{s}\in\mathcal{V'}$ that defines $\mathcal{V}_{e}$ set that splits $\mathcal{G}$ into two parts. 
The partitioning layer $v_{s}$ is the last one to be processed in the edge. Hence, the layers from $v_{1}$ to $v_{s}$ are processed at edge. Then, the edge device sends the output data of $v_{s}$ to the cloud that processes the next layers.

Once the partitioning layer $v_{s}$ is found, we can determine which set each layer belongs to.
Formally, the set processed at the edge is $\mathcal{V}_{e} = \mathcal{V'}\cup\mathcal{B}$,
where $\mathcal{V'} = \{v_{1}, \cdots, v_{s}\}$ and $\mathcal{B} = \{b_{1}, \cdots, b_{s-1}\}$. Hence, the set processed in the cloud is $\mathcal{V}_{c} = \{v_{s+1}, \cdots, v_{|\mathcal{V}|}\}$. 
It is important to note, in this paper, that no side branch is processed in the cloud, hence no vertex $b_{i}\in\mathcal{B}$ belongs to the set $\mathcal{V}_{c}$. Therefore, all vertices $b_{i}$ posterior to $v_{s}$ are discarded. It occurs because, in the cloud, the reduction of processing delays provided by classifying a sample at the side branch is negligible when compared with the time required to execute all layers of the main branch. Moreover, this reduction is significantly lower than the communication delay between the edge and the cloud. 
Figure~\ref{fig:partitioning_scenarios} shows different partitioning scenarios for a BranchyNet composed of four layers.

In Figure~\ref{fig:partitioning_scenarios}, the gray vertices represent the layers processed at the edge device, the orange vertex corresponds to the partitioning layer $v_{s}$ and the blue vertices refer to layers processed at the cloud server. 
Figure~\ref{fig:only_edge_processing} illustrates an example where all layers are processed in the edge, so the partitioning layer is $v_{4}$. Thus, we have $\mathcal{V}_{e} = \{v_{1},v_{2},v_{3},v_{4}, b_{1},b_{2},b_{3}\}$ and $\mathcal{V}_{c} = \emptyset$. In this case, no data is sent to the cloud. 
On the other hand, Figure~\ref{fig:only_cloud_processing} shows cloud-only processing where all layers are processed at the cloud, so $\mathcal{V}_{e} = \emptyset$ and $\mathcal{V}_{c} = \{v_{1},v_{2},v_{3},v_{4}\}$. 
In this case, the output data size sent to the cloud corresponds to the raw input data size. 
Finally, Figure~\ref{fig:partitioning_processing} shows an example where partition occurs in the layer $v_{2}$, so $\mathcal{V}_{e} = \{v_{1},b_{1},v_{2}\}$ and $\mathcal{V}_{c} = \{v_{3},v_{4}\}$. 
In this case, the output data of layer $v_{2}$ is sent from an edge device to the cloud. 

Generally, the partitioning splits a BranchyNet to follow an objective, such as reducing inference time, saving bandwidth, maximizing inference accuracy, or even reducing energy consumption. In this paper, our goal is to minimize the inference time. We derive next a model to estimate the inference time.

\subsection{Estimation of Inference Time}
\label{subsec:inference_time}
In DNN applications with edge computing, end-devices send input data to the edge device. When the DNN has no side branches, this data is processed by all layers placed at the edge.
The edge device sends the output data of the partitioning layer $v_{s}$ to the cloud, which is responsible to process the remaining layers. 
Hence, in this DNN, the inference time depends on the processing delay in the edge and the cloud, as well as their communication delay. 

Edge devices and cloud servers differ significantly regarding their computational power and thus have different processing delays. Hence, the processing time for a given layer depends on where it is computed.
Let $t_{i}^{e}$ and $t_{i}^{c}$ be the processing time to compute the layer $v_{i}$ at the edge and the cloud, respectively.
The total processing delay at the edge, when the DNN has no side branches, is 
\begin{equation}
T_{e} = \sum_{i|v_{i}\in\mathcal{V}_{e}} t_{i}^{e},
\label{eq:dnn_edge_processing_time}
\end{equation}
and the processing time to compute all the remaining layers at cloud is given by 
\begin{equation}
T_{c} = \sum_{i|v_{i}\in\mathcal{V}_{c}} t_{i}^{c}.
\label{eq:dnn_cloud_processing_time}
\end{equation} 
The communication delay depends on the output data size of the partitioning layer and the network bandwidth. 
The output data size generated by each DNN layer presents a non-monotonic behavior, which means each layer produces output data with different sizes.
Therefore, there are layers closer to the input layer
that generate less data than a deeper layer (i.e., closer to output layer), resulting in higher communication delay.
For each layer $v_{i}$, we can define the communication time as $t_{i}^{\text{net}} = \frac{\alpha_{i}}{B}$
where $\alpha_{i}$ is the output data size of layer $v_{i}$ and $B$ is the network bandwidth. 
At this stage, we can define $\boldsymbol{t_{e}} = [t_{1}^{e}, \cdots, t_{|\mathcal{V}|}^{e}]$ and 
$\boldsymbol{t_{c}} = [t_{1}^{c}, \cdots, t_{|\mathcal{V}|}^{c}]$ as parameters related to 
hardware resources and $\boldsymbol{t_{\text{net}}} = [t_{1}^{\text{net}}, \cdots, t_{|\mathcal{V}|}^{\text{net}}]$ related to network condition. 
Additionally, we can assign a 3-tuple  $(t_{i}^{e},t_{i}^{\text{net}},t_{i}^{c})$ to each vertex $v_{i}$ of a DNN with no side branches or for the main branch in a BranchyNet $\mathcal{G}_{\text{BDNN}}$.

In a DNN with no side branches, the inference time is the sum of the total processing delay (i.e. $T_{e}$ and $T_{c}$) with the communication delay (i.e., $t_{s}^{\text{net}}$), as shown in Equation~\ref{eq:dnn_inference_time}~\cite{hu2019dynamic}.
\begin{equation}
T_{inf}^{DNN} = T_{e} + t_{s}^{\text{net}} + T_{c}
\label{eq:dnn_inference_time}
\end{equation}

If we consider a DNN with side branches (i.e., a BranchyNet), an inference can stop at one of these branches, if the classification achieves a certain confidence criterion. To model the inference time for DNN with side branches, we divide our analysis into a particular case and a generic one.
\subsubsection{Particular Case} 
\label{subsub:particular_case}
We first consider a particular BranchyNet, with only one side branch $b_{k}$, placed in the output of any middle layer $v_{k}$, where $1 \leq k \leq (|\mathcal{V}| - 1)$. As BranchyNet inference algorithm described in Section~\ref{sec:branchyNet}, in this case, when a sample reaches this side branch $b_{k}$, it can be classified and exits the BranchyNet at that middle layer or is processed by the next layers until reach the output layer. To model this two possible outcomes, we define a Bernoulli random variable $X_{k}$ that takes on 1 if a sample is classified at the side branch $b_{k}$ with probability $\prob[X_{k}=1] = p_{k}$, where $p_{k}\in[0,1]$. Otherwise, $X_{k}=0$ if the sample is process 
by the next layers to side branch $b_{k}$ with $\prob[X_{k} = 0] = 1 - p_{k}$.
 
\subsubsection{General Case} 
\label{subsub:general_case}
We generalize the previous analysis, considering a BranchyNet with $|\mathcal{V}|- 1$ side branches, as shown in Figure~\ref{fig:branchyNet}.
As in particular case, we define a Bernoulli random variable $X_{k}$ for each side branch $b_{k}\in\mathcal{B}$, resulting in a sequence of Bernoulli random variables $\boldsymbol{X}=(X_{1},X_{2},\cdots,X_{|\mathcal{V}|-1})$.
For a sample to be classified at side branch $b_{k}$, this sample cannot meet the confidence criterion in any of the $k-1$ previous side branches. 
Thus, 
the Bernoulli random variable $X_{k}$ takes on value 1 for the first time at side branch $b_{k}$ after $k-1$ trials. To model it, we define a random variable $Y$ that represents the number of side branches, which have already been processed before the current side branch $b_{k}$, and it cannot classify the sample. Then, the probability of random variable Y as follows:  
\begin{equation}
    p_{Y}(k)= \prob[Y = k] = p_{k}\prod_{i=1}^{k-1}(1-p_{i}).
\label{eq:geometric_distribution}    
\end{equation}

Considering a batch of input data samples, we can compute the expected value of the inference time. As the cloud has no side branches, if the partitioning occurs earlier than the first side branch, the inference time is modeled by Equation~\ref{eq:dnn_inference_time} as a DNN with no side branches. 
If the partitioning occurs after the side branch, the sample can be classified by this branch, and thus it is not processed by the remaining layers. 
In this case, in a BranchyNet with only one side branch, the expected value of the inference time is
\begin{equation}
\EX[T_{inf}^{\text{BDNN}}(k)] = \sum_{\mathclap{\substack{i|v_{i}\in\mathcal{V}_{e}\\i\leq k}}}t_{i}^{e} + (1-p_{Y}(k))\left(\sum_{\mathclap{\substack{\;\,\,i|v_{i}\in\mathcal{V}_{e}\\i\leq k}}}t_{i}^{e} + t_{s}^{\text{net}} + T_{c}\right).
\label{eq:branchyNet_value_expected_inference_time}    
\end{equation}

Equation~\ref{eq:branchyNet_value_expected_inference_time} 
shows that the edge device always processes the layers before the side branch $b_{k}$. However, the processing and communication delays of any remaining layers are weighted by the probability of classifying the input data at side branch.
In an extreme case, where the input samples are always classified at the side branch, which means $p=1$, Equation~\ref{eq:branchyNet_value_expected_inference_time} considers neither the communication delay nor the processing delay for the remaining layers. On the other extreme, if the inference never stops at a side branch (i.e., $p=0$), Equation~\ref{eq:branchyNet_value_expected_inference_time} is equal to Equation~\ref{eq:dnn_inference_time}.
At this stage, according to the partitioning layer position, the expected value of the inference time can be modeled as follows:
\begin{equation}
\EX[T_{inf}(k)] = \left\{
\begin{array}{lll}
T_{inf}^{\text{DNN}}, & \text{$\forall v_{k}$, $v_{s}$ $|$ $s < k$.}\\
\EX[T_{inf}^{\text{BDNN}}], &\text{otherwise.}
\end{array}
\right.
\label{eq:final_inference_time_two_cases}
\end{equation}

\section{BranchyNet Partitioning Optimization}
\label{sec:optimization_formulation}
\begin{figure}[!ht]
    \centering
    \includegraphics[width=.9\linewidth]{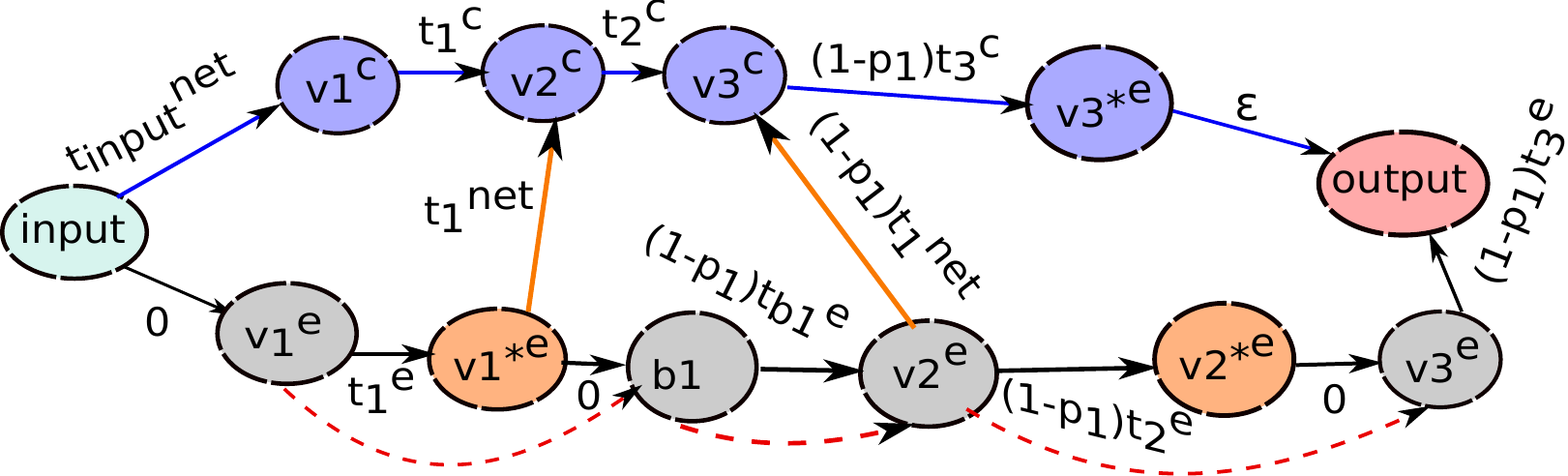}
    \caption{Graph representation $\mathcal{G'}_{\text{BDNN}}$ of 3-layer BranchyNet.}
    \label{fig:shortest_path_model}
\end{figure}

In this section, our goal is to determine the partitions $\mathcal{V}_{e}$, $\mathcal{V}_{c}$ that minimize the inference time of Equation~\ref{eq:final_inference_time_two_cases}, given the input parameters $\boldsymbol{t_{e}}$, $\boldsymbol{t_{c}}$, $\boldsymbol{t_{\text{net}}}$, $B$ and $p_{Y}(k)$. As defined in Section~\ref{subsec:graph_branchyNet}, the main branch of BranchyNet is modeled as a chain graph $\mathcal{P}_{|\mathcal{V'}|}$. Hence, if we find a partitioning layer $v_{s}$ that minimize $\EX[T_{inf}(k)]$, we can determine the partitions $\mathcal{V}_{e}$ and $\mathcal{V}_{c}$.
To this end, we propose to construct a new graph $\mathcal{G'}_{\text{BDNN}}$ based on the BranchyNet graph. These graph $\mathcal{G'}_{\text{BDNN}}$ allows to associate one delay of the 
a 3-tuple $(t_{i}^{e}$, $t_{i}^{c}$, $t_{i}^{\text{net}})$ to each link, as described Section~\ref{subsec:inference_time}.  
The delay associated with each link $(v_{i}, v_{j})$ depends on where the vertex $v_{i}$ is processed. 
When then we show that the partitioning problem can be considered as the shortest path problem in $\mathcal{G'}_{\text{BDNN}}$. 
To build $\mathcal{G'}_{\text{BDNN}}$, we create two disjoint chain graphs $\mathcal{P}_{|\mathcal{V'}\cup\mathcal{B}|}^{e}$ and $\mathcal{P}_{|\mathcal{V'}|}^{c}$, such as those of Figures~\ref{fig:only_edge_processing} and~\ref{fig:only_cloud_processing}, respectively. 
The vertices of $\mathcal{P}_{|\mathcal{V'}\cup\mathcal{B}|}^{e}$ and $\mathcal{P}_{|\mathcal{V'}|}^{c}$ represent the layers processed in the edge in the cloud, respectively. Then, we assign a weight $\omega_{(v_{i}, v_{j})}$ to each link $(v_{i}, v_{j})$ of the graphs $\mathcal{P}_{|\mathcal{V'}\cup\mathcal{B}|}^{e}$ and $\mathcal{P}_{|\mathcal{V'}|}^{c}$, representing the processing time to compute $v_{i}$ at the edge (i.e., $t_{i}^{e}$) and at cloud (i.e., $t_{i}^{c}$), respectively.
Figure~\ref{fig:shortest_path_model} shows the graph $\mathcal{G'}_{\text{BDNN}}$ constructed based on a BranchyNet constituted by the main branch with three layers and one side branch inserted after the first layer of the main branch. 
In this figure, each gray and blue vertices represents a layer processed at the edge and cloud, respectively. The dashed red and blue links correspond to links of graphs $\mathcal{P}_{|\mathcal{V'}\cup\mathcal{B}|}^{e}$ and $\mathcal{P}_{|\mathcal{V'}|}^{c}$. 

The graph $\mathcal{G'}_{\text{BDNN}}$ must model three possibilities: edge-only processing, cloud-only processing, and processing with partitioning. 
To model cloud-only and edge-only, we introduce two virtual vertices called $input$ and $output$ and then, we add links $(\text{input},v_{1}^{c})$ and $(\text{input},v_{1}^{e})$ into graph $\mathcal{G'}_{\text{BDNN}}$, illustrated by a blue and black links in Figure~\ref{fig:shortest_path_model}. Then, we assign weights to these links that are related to communication delay in cloud-only and edge-only processing. The weight $\omega_{(\text{input}, v_{i}^{c})}$ corresponds to communication time to upload a raw input sample to the cloud denoted by $t_{\text{input}}^{\text{net}}$ in cloud-only processing. In edge-only, the weight $\omega_{(\text{input}, v_{i}^{e})}=0$ since there is not communication delay. Figure~\ref{fig:shortest_path_model} shows that the path between $input$ and $output$ using only the red dashed and blue links computes the inference time for edge-only and cloud-only processing, respectively.  
To model the processing with partitioning in graph $\mathcal{G'}_{\text{BDNN}}$, for all $v_{i}^{e}\in\mathcal{V'}$, we introduce an auxiliary vertex  $v_{i}^{*e}\in\mathcal{V}_{A}$ (i.e., orange vertices in Figure~\ref{fig:shortest_path_model}), where $\mathcal{V}_{A}$ is the set of auxiliary vertices. Then, we add a link $(v_{i}^{e},v_{i}^{*e})$  and replace $(v_{i}^{e},v_{i+1}^{e})$ (i.e., red dashed links in Figure~\ref{fig:shortest_path_model}) to $(v_{i}^{*e},v_{i+1}^{e})$. In Figure~\ref{fig:shortest_path_model}, the red dashed links $(v_{1}^{e}, b_{1})$ are replaced to the black links $(v_{1}^{e}, v_{1}^{*e})$ and $(v_{1}^{*e}, b_{1})$. 
To model the communication between edge and cloud, we add a link, whose the weight corresponds to the communication time to send output data of partitioning layer $v_{s}\in\mathcal{V'}$ placed at edge to $v_{s+1}\in\mathcal{V}_{c}$ placed at cloud denoted by $t_{s}^{\text{net}}$. 
In this figure, the orange link, such as $(v_{1}^{*e}, v_{2}^{e})$, represents the communication between edge side and cloud. 
To avoid ambiguity in the choice of the shortest path when the probability $p_{Y}(k) = 1$, we add a virtual vertex $v_{3}^{*c}$ as successor of $v_{3}^{c}$ and predecessor of vertex $output$. Then, we assign the weight $\epsilon$ to the link $(v_{3}^{*c}, \text{output})$. The weight $\epsilon$ must be a very small value, to not interfere with the result of the shortest path problem. If the probability $p_{Y}(k) = 0$, which means the probability that any sample is classified by side branch $b_{1}$, the graph $\mathcal{G'}_{\text{BDNN}}$ represents a regular DNN.  
The weights are assigned to links of $\mathcal{G'}_{\text{BDNN}}$ as follows:
\begin{equation}
\label{eq:weights_links}
\omega_{(v_{i}, v_{j})} = \left\{
\begin{array}{lll}
t_{i}^{e}, & \text{if $v_{i}\in\mathcal{V}_{e}$}\\
t_{i}^{c}, & \text{if $v_{i}\in\mathcal{V}_{c}$}\\
t_{i}^{\text{net}}, & \text{if $(v_{i}\in\mathcal{V}_{A}$, $v_{j}\in\mathcal{V}_{c})\vee(v_{i}=\text{input})$}\\
\epsilon, & \text{if $v_{i}\in\mathcal{V}_{c}$, $v_{j} = \text{output}$, }\\
0, & \text{if $v_{i}\in\mathcal{V}_{A}$, $v_{j}\in\mathcal{V}_{e}$}
\end{array}
\right.
\end{equation}

To model the expected value of inference time in a BranchyNet, the weights $\omega_{(v_{i}, v_{j})}^{\text{BDNN}}$ assigned to the links in $\mathcal{G'}_{\text{BDNN}}$ is weighted by the probability $p_{Y}(k)$ that the sample is classified at side branch $b_{k}$. Thereby, as higher the probability that a sample is classified at the side branch, less significant are the weights of links after the side branch. Thus, the weights assigned to links in a Branchynet is given by 
\begin{equation}
 \omega_{(v_{i}, v_{j})}^{\text{BDNN}}(k) = p_{Y}(k)\omega_{(v_{i}, v_{j})}.   
\end{equation}

Our goal is to determine the path with the minimum cost that connects the virtual vertices $input$ and $output$. The cost of the path is defined as the sum of the weights associated with the links in graph $\mathcal{G'}_{\text{BDNN}}$. 
At this point, we show the equivalency between the BranchyNet partitioning problem and the shortest path problem.
Given two vertices, the shortest path problem finds a forward path that connects these two vertices with minimum cost. In this problem, these two vertices are $input$ and $output$.
In Figure~\ref{fig:shortest_path_model},
if the layers of $\mathcal{V}_{e}$ are contained in the shortest path, it means that processing strategy is edge-only. 
In this case, the total cost correspondents to $T_{e}$, as shown in Equation~\ref{eq:dnn_edge_processing_time}. 
However, if all vertices $v_{i}\in\mathcal{V}_{c}$ is contained in the shortest path, it means cloud-only processing.
The cost of shortest path is thus $T_{c}$, as defined in Equation~\ref{eq:dnn_cloud_processing_time}. Otherwise, if the vertices in shortest path belongs to $\mathcal{V}_{e}$ as well as $\mathcal{V}_{c}$, then partitioning occurs. The total cost is given by Equation~\ref{eq:final_inference_time_two_cases}. For instance, in Figure~\ref{fig:shortest_path_model}, assuming the partitioning layer is vertex $v_{2}$, the shortest path is $(\text{input} ,v_{1}^{e})$, $(v_{1}^{e}, v'^{e}_{1})$, $(v'^{e}_{1}, b_{1})$, $(b_{1}, v_{2}^{e})$, $(v_{2}^{e}, v'^{e}_{2})$, $(v^{e}_{2}, v_{3}^{c})$, $(v_{3}^{c}, v_{3}^{*c})$ and $(v_{3}^{*c}, \text{output})$. 
Thus, the layers $v_{1}$ through $v_{2}$ are processed at the edge and belong to set $\mathcal{V}_{e}$ then the edge sends output data of $v_{2}$ to the cloud, which, in turn, processes the $v_{3}$ that belongs to set $\mathcal{V}_{c}$. 

The shortest path problem is a well-known problem that can be solved in polynomial-time. In this work, Dijkstra's algorithm is used to find the shortest path with the computational complexity of $\mathcal{O}(m + n\log(n))$, where $m$ and $n$ are the number of links and vertices in $\mathcal{G'}_{\text{BDNN}}$, respectively.

\section{Evaluation}
\label{sec:experiments}
\begin{figure*}[ht!]
\subfigure[Processing factor of 10.]{\label{fig:probability_inference_time_factor_10}\includegraphics[width=0.3\linewidth]{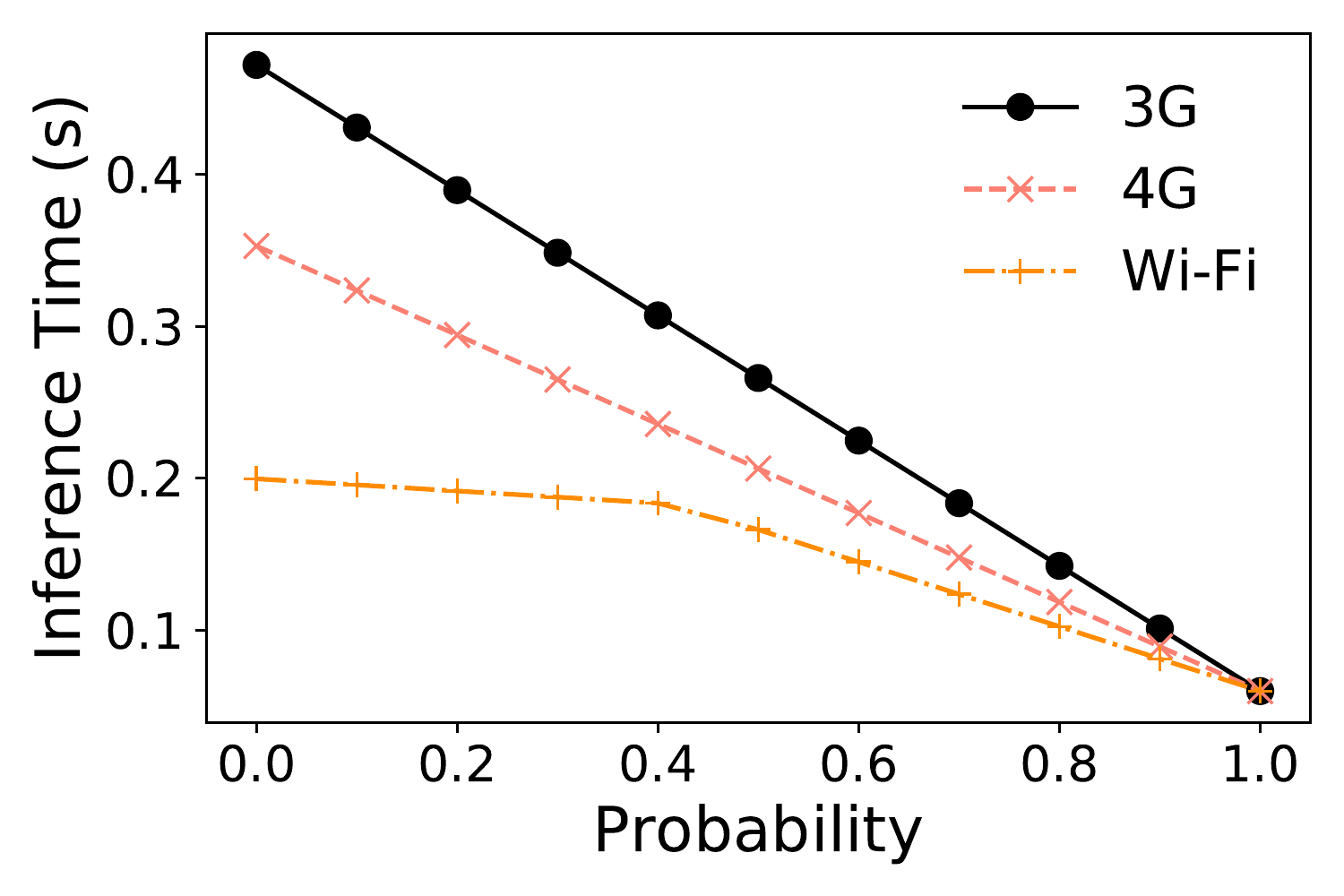}}
\hspace{5mm}
\subfigure[Processing factor of 100.]{\label{fig:probability_inference_time_factor_100}\includegraphics[width=0.3\linewidth]{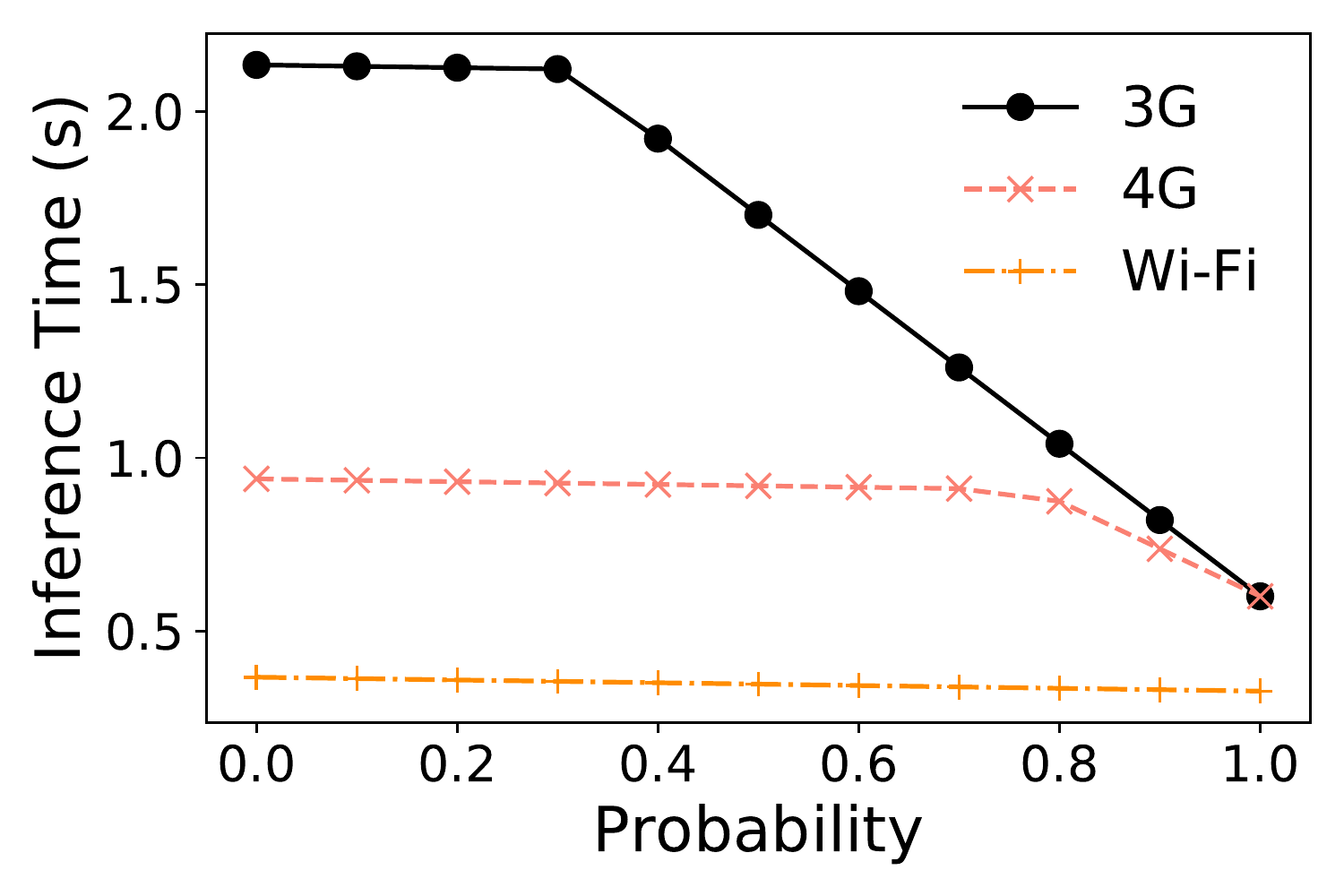}}
\hspace{5mm}
\subfigure[Processing factor of 1000.]{\label{fig:probability_inference_time_factor_1000}\includegraphics[width=0.3\linewidth]{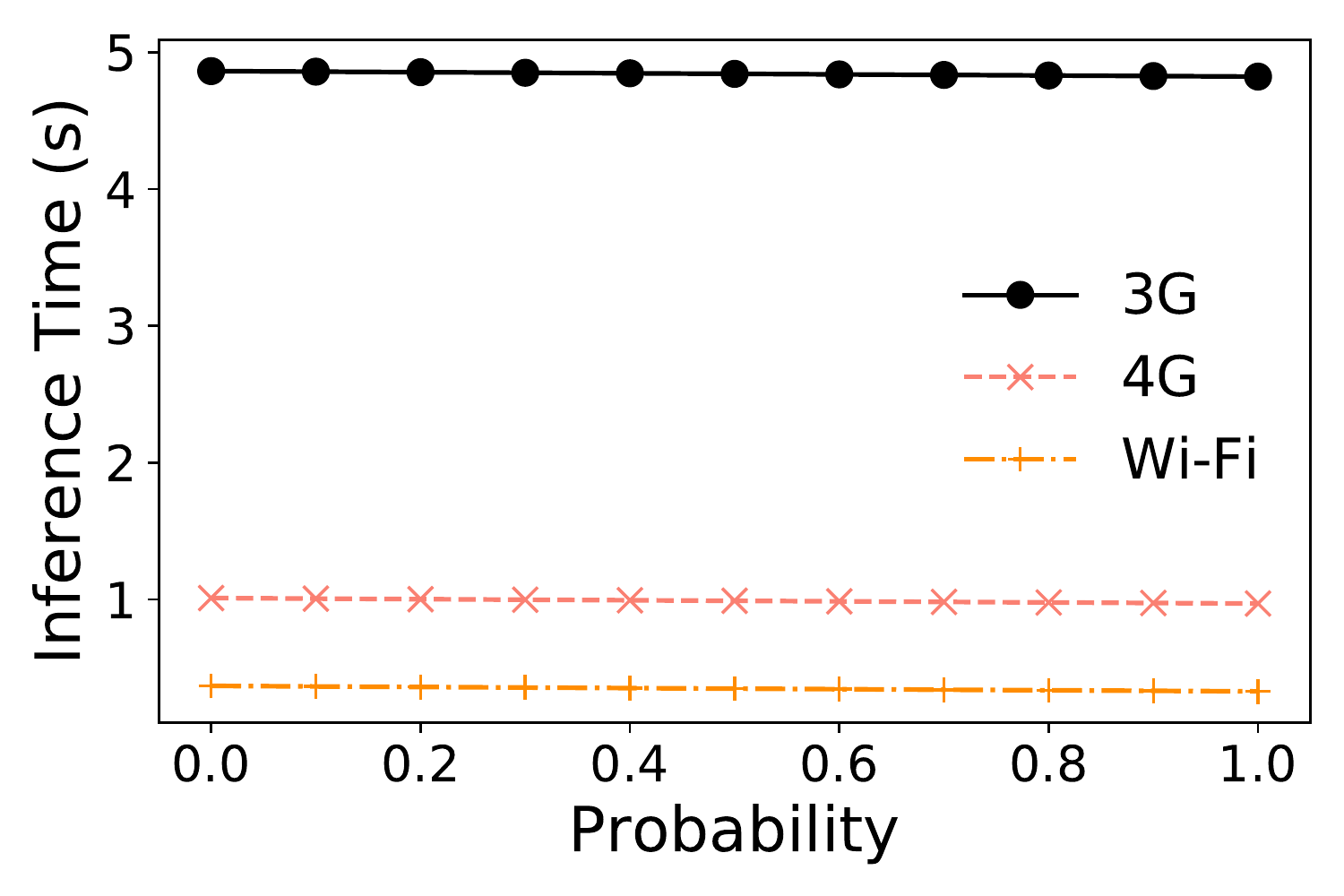}}
\caption{Inference time according to the probability of classifying a sample for different wireless technology and processing factors $\gamma$}
\label{fig:probability_inference_time}
\end{figure*}
The experiments present a sensitivity analysis evaluating the impacts of input parameters described in Section~\ref{subsec:inference_time}, such as $t_{i}^{c}$, $t_{i}^{e}$, $B$ and $p_{Y}(k)$, in inference time, under different network bandwidth. Then, we analyze the partitioning layer choice under different processing capacities of the cloud server and edge device.
To this end, we implement a BranchyNet called B-AlexNet~\cite{teerapittayanon2016branchynet}. The B-AlexNet is composed of a standard AlexNet architecture as the main branch
with one side branch inserted after the first middle layer of the main branch. The choice of only one side branch aims to simplify the experimental analysis, and the side branch position is chosen to avoid unnecessary processing in the edge. First, we have to obtain those input parameters.
The parameters $\mathcal{G'}_{\text{BDNN}}$ and the output data size $\alpha_{i}$ for each layer are obtained after the definition of B-AlexNet as BranchyNet architecture. 
To obtain the communication time $t_{i}^{\text{net}}$, we consider that the edge device uses wireless technology to send data to the cloud. Besides, we assume that the bottleneck is the access network. We thus use average uplink rates of 1.10, 5.85, 18.80\,Mbps, which corresponds to 3G, 4G, and Wi-Fi, respectively. 
These average uplink rates used in this work are based on the values presented in~\cite{hu2019dynamic}.
To obtain $\boldsymbol{t_{c}}$, we measure the processing time for each layer of B-AlexNet, using Google Colaboratory.
Google Colaboratory is a cloud computing service for machine learning, equipped with Intel 2-core Xeon(R)@ 2.20GHz processor, 12 GB VRAM and GPU Tesla K80.
We define the processing time at edge for each layer as $t_{i}^{e} = \gamma \cdot t_{i}^{c}$
where $\gamma\in\mathbb{N}_{>1}$ is a proportionality factor that indicates the ratio between processing time at cloud and edge. 
This factor spans different edge hardware types in our evaluation. For example, Jetson TX2
module developed by NVIDIA can represent an edge device equipped with high processing power and thus a low $\gamma$. On the other hand, Raspberry Pi
can represent a resource-constraint device, with a high $\gamma$. 

Figure~\ref{fig:probability_inference_time} shows the impact of probability in expected value of inference time, given by Equation~\ref{eq:branchyNet_value_expected_inference_time}, for three processing factors: 10, 100 and 1000. These results are obtained based on the solution of our optimization problem when varying the probability that a sample is classified at the side branch. For each processing factor, we present results for 3G, 4G and Wi-Fi. Figure~\ref{fig:probability_inference_time} shows that, when the edge has high processing power, the probability has a severe impact on inference time. Also note that, for each processing time, the y-axis has a different scale, meaning that a high edge processing power leads to an overall low inference time. 

Considering a given processing factor, Figures~\ref{fig:probability_inference_time_factor_10}~and~\ref{fig:probability_inference_time_factor_100} show that networks with lower bandwidth are more affected by probability. For example, the 3G results in Figure~\ref{fig:probability_inference_time_factor_10} show that the inference time reduces 87.27\% if we compare a case where the probability is zero with the case where the probability is one. On the other hand, this difference is 82.98\% and 70\%, for  4G and Wi-Fi, respectively. Figure~\ref{fig:probability_inference_time_factor_10} also shows that, when the probability is one, all network technologies have the same inference time. In this case, this result is expected since all samples are classified at the side branch. 
Although the probability also affects inference time in Figure~\ref{fig:probability_inference_time_factor_100}, we can note that, for each technology, the y-axis remains constant for a given range of probability. This explained since, as compared to the case of Figure~\ref{fig:probability_inference_time_factor_10}, the edge has lower processing power. Hence, for low probabilities, the optimization problem chooses cloud-only processing since the major part of the samples are not classified on the side branch. As cloud-only processing does not have side branches, the inference time is not affected by the probability.
After a given probability value, the problem begins to choose partitioning solutions, where the edge is involved, and thus the probability begins to affect the inference time. For example, for 3G, the inference time only starts to decrease when the probability is higher than 0.3. For 4G, this value is 0.8 since, as compared to 3G, the bandwidth is higher, and thus it is more interesting to send raw data to the cloud for a large probability range. Finally, the Wi-Fi results of Figure~\ref{fig:probability_inference_time_factor_100} shows that it is always interesting to perform cloud-only processing due to its high bandwidth.
Figure~\ref{fig:probability_inference_time_factor_100} shows an extreme situation where the probability does not affect the inference time. This behavior happens because the edge has low processing power, and thus it is always interesting to perform cloud-only processing.

\begin{figure}[ht!]
\centering
\subfigure[Partitioning Layer according to factor $\gamma$ with 3G.]{\label{fig:processing_X_partitioning_1}\includegraphics[width=0.7\linewidth]{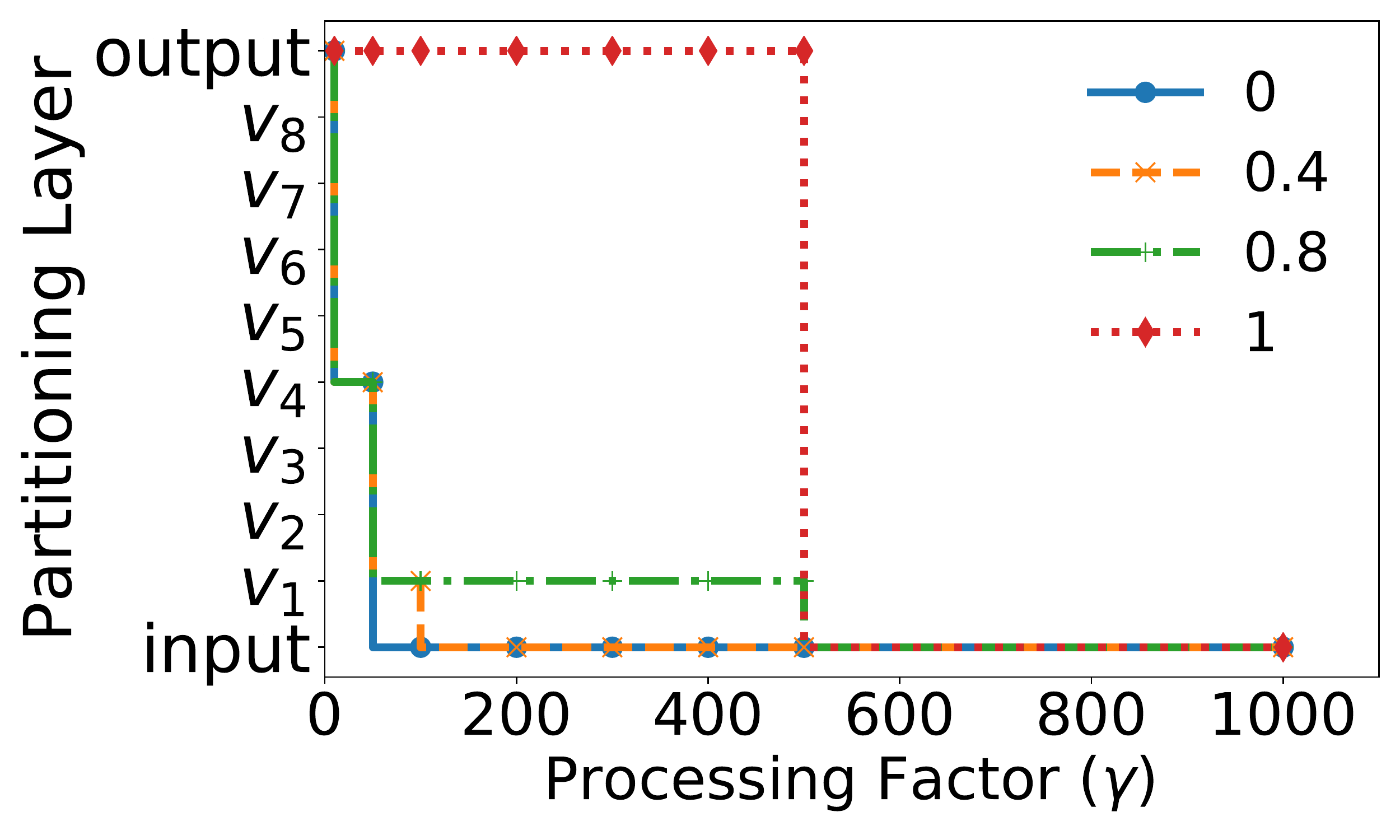}}
\hspace{5mm}
\subfigure[Partitioning Layer according to factor $\gamma$ with 4G.]{\label{fig:processing_X_partitioning_5}\includegraphics[width=0.7\linewidth]{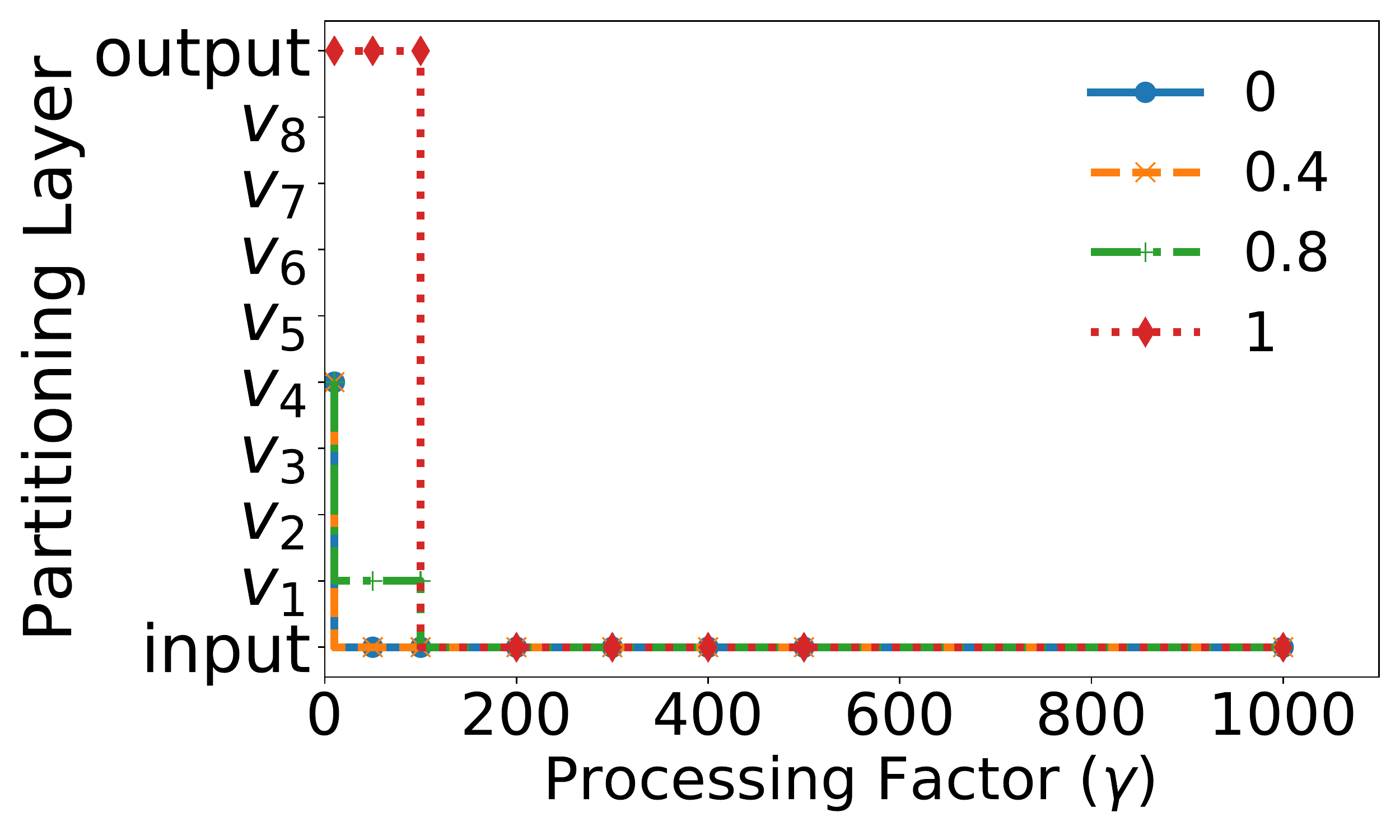}}
\caption{Partitioning layer for different processing factors.}
\label{fig:partLayerEva}
\end{figure}

Using the same scenario of Figure~\ref{fig:probability_inference_time}, we vary the processing factor $\gamma$ and analyze which layer the optimization problem chooses as the partitioning one. Figure~\ref{fig:processing_X_partitioning_1} and~\ref{fig:processing_X_partitioning_5} shows the chosen layer for different processing factors, when using 3G and 4G, respectively. Each curve represents a given probability of classifying the sample at a side branch. This behavior is expected since an edge with a lower processing power means that it is more interesting to process the layers in the cloud.
Figure~\ref{fig:processing_X_partitioning_1} shows that as the processing factor increases, the chosen partitioning layer moves toward to input layer.
For example, assuming a probability $p = 0.8$, when $\gamma$ changes from 500 to 600, Figure~\ref{fig:processing_X_partitioning_1} shows that the partitioning layer changes from $conv1$ to $input$, which means cloud-only processing. 
When comparing Figure~\ref{fig:processing_X_partitioning_5} with Figure~\ref{fig:processing_X_partitioning_1}, we note that, for 4G, the problem starts to choose cloud-only processing for a lower processing factor (i.e., a higher edge processing power). This confirms the trend observed in Figure~\ref{fig:probability_inference_time}, where, for a higher average uplink rate, the problems tend to choose cloud-only processing.
Finally, Figure~\ref{fig:partLayerEva} also confirms the behavior of Figure~\ref{fig:probability_inference_time} where the probability affects the choice of the partitioning layer, thus impacting the inference time. 

\begin{figure}[!ht]
    \centering
    \includegraphics[width=0.7\linewidth]{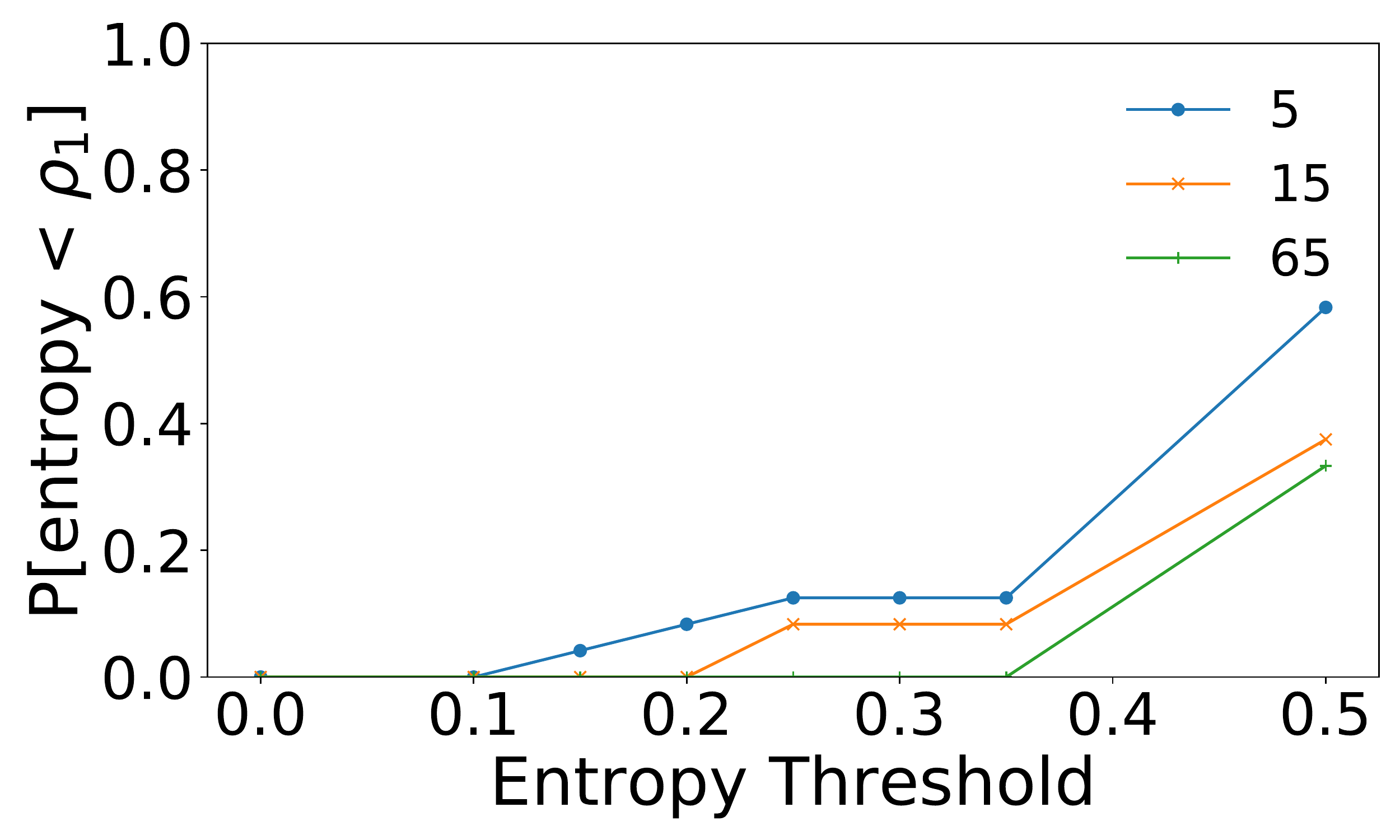}
    \caption{Probability of side branch classification under different distortion levels in B-AlexNet.}
    \label{fig:entropy_threshold}
\end{figure}

In practice, the probability of classifying an input image at a side branch is a parameter related to aspects inherent of input data that depends on numerous factors. 
One of these factors is image quality. To show that image quality affects the probability, and thus the partitioning decision, we train a B-AlexNet for image classification, using a cat-and-dog dataset~\cite{dodge2016understanding}. This dataset is composed of images of dogs and cats in different environments without any distortion. Once trained, we apply a batch with 48 samples with different levels of Gaussian blur. 
This experiment implements Gaussian filters with dimensions 5, 15, 65 to represent images with low, intermediate, and high distortion, respectively. 
We use these images as samples to perform inferences to classify if an animal is a dog or a cat.
Figure~\ref{fig:entropy_threshold} shows the probability of classifying an input image according to the entropy threshold.
This figure shows that as distortion level increases, the probability that a sample is classified at a side branch decreases. This is true since images with higher distortion levels tend to have a higher uncertainty in the inference, resulting in a lower probability that it is classified at a side branch.

\section{Conclusion}
\label{sec:conclusion}
In this paper, we accelerate DNN inference, partitioning DNN between the edge device and cloud server, to minimize the inference time. Different from a traditional DNN, the BranchyNet has side branches that allow the inference to stop at the middle layers, which can reduce the inference time. Hence, to partition a BranchyNet we have to take into account the probability that inferences stop at a side branch. To address this problem, we model the expected value of inference time as a function of different factors, such as processing and communication delay and the probability of classification at a side branch. We also model the BranchyNet as a graph, showing that the minimization of inference time can be solved as the shortest path problem. Hence, this problem can be solved in polynomial-time, using Dijkstra's algorithm. We evaluate our problem using a sensitivity analysis where we vary the probability of side branch classification and the processing power in the edge. 
From the results obtained, we demonstrate that probability 
affects partitioning layer choice, hence impacts the inference time. 
Thus, estimating the probability allows improving the partitioning decision as network conditions
and computational resources.
Moreover, evaluations show that processing strategy changes according to the probability. Thus, this paper also introduces the probability as a factor to be considered in BranchyNet partitioning.  
As future work, our first goal is to extend our proposal to handle also DAG topology DNN. 
Moreover, we will investigate heuristics for side branch placement, to attempt also accuracy requirement.

\IEEEpeerreviewmaketitle

\section*{Acknowledgement}
This study was financed in part by the Coordenação de Aperfeiçoamento de Pessoal de Nível Superior - Brasil (CAPES) - Finance Code 001. It was also supported by CNPq, FAPERJ Grants E-26/203.211/2017, E-26/201.833/2019, and E-26/010.002174/2019, and FAPESP Grant 15/24494-8.

\bibliographystyle{IEEEtran}
\bibliography{header,dcn}

\end{document}